\documentstyle[12pt,epsfig]{article}
\renewcommand{\baselinestretch}{1.1}
\newcommand{\beq}{\begin{equation}}
\newcommand{\eeq}{\end{equation}}
\newcommand{\api}{\frac{\alpha_s}{\pi}}
\newcommand{\ba}{\begin{array}}
\newcommand{\ea}{\end{array}}

\newcommand{\msbar}{\overline{\mbox{MS}}}
\newcommand{\dsp}{\displaystyle}
\newcommand{\EQN}{\label}

\begin{document}

\begin{titlepage}
%
%
\mbox{}
\hfill LBNL--37881 \\
\mbox{}
\hfill TTP 95--35\footnote{The complete postscript file of this
preprint, including figures, is available via anonymous ftp at
ttpux2.physik.uni-karlsruhe.de (129.13.102.139) as
/ttp95-35/ttp95-35.ps or via www at
http://ttpux2.physik.uni-karlsruhe.de/cgi-bin/preprints/
Report-no: TTP95-35 and at
http://theor1.lbl.gov/www/theorgroup/papers/37881.ps.} \\
\mbox{}
\hfill hep-ph/9510456\\
\mbox{}
\hfill October 1995

\protect\vspace*{.8cm}
%
%
\begin{center}
  \begin{Large}
  \begin{bf}
The $\msbar$ Renormalized Bottom Mass of Order
${\cal O}(\alpha_s G_F M_t^2)$ and its Application to
$\Gamma(H\to b\bar{b})$\footnote{
                            This work was in part supported by
                            US DoE under Contract DE-AC03-76SF00098.
                                }
\\
  \end{bf}
  \end{Large}
%
%
  \vspace{0.3cm}
A.~Kwiatkowski$^{a}$\footnote{
                           Supported by Deutsche Forschungsgemeinschaft,
                           grant no. Kw 8/1-1.}
M.~Steinhauser$^{b}$
%

\begin{itemize}
\item[$^a$]
              Theoretical Physics Group \\
              Lawrence Berkeley National Laboratory\\
              University of California\\
              Berkeley, CA. 94720, USA
\item[$^b$]
    Institut f\"ur Theoretische Teilchenphysik\\
    Universit\"at Karlsruhe \\
    D-76128 Karlsruhe, Germany
\end{itemize}
\vspace{0.2cm}

%
%
  \vspace{0.5cm}
  {\bf Abstract}
\end{center}
\begin{quotation}
\noindent
The renormalized mass of the bottom quark is
calculated at the two loop level to order
${\cal O}(\alpha_s G_F M_t^2)$ in the $\msbar$
renormalization scheme. Different strategies for
the computation are outlined.
 The result is
applied to the partial decay rate
 $\Gamma(H\rightarrow b\bar{b})$ of the Higgs boson
 into bottom quarks.
Expressing the width in
terms of the running mass instead of the bottom
pole mass allows to treat the  ${\cal O}(\alpha_s G_F M_t^2)$
radiative corrections  on the same footing
 as is commonly used in pure QCD calculations.
The numerical values for the corrections
  are given and the sizes of
 different contributions are compared.
\end{quotation}

\vfill

%

\end{titlepage}

\renewcommand{\thepage}{\roman{page}}
\setcounter{page}{2}
\mbox{ }

\vskip 1in

\begin{center}
{\bf Disclaimer}
\end{center}

\vskip .2in

\begin{scriptsize}
\begin{quotation}
This document was prepared as an account of work sponsored by the United
States Government. While this document is believed to contain correct
 information, neither the United States Government nor any agency
thereof, nor The Regents of the University of California, nor any of their
employees, makes any warranty, express or implied, or assumes any legal
liability or responsibility for the accuracy, completeness, or usefulness
of any information, apparatus, product, or process disclosed, or represents
that its use would not infringe privately owned rights.  Reference herein
to any specific commercial products process, or service by its trade name,
trademark, manufacturer, or otherwise, does not necessarily constitute or
imply its endorsement, recommendation, or favoring by the United States
Government or any agency thereof, or The Regents of the University of
California.  The views and opinions of authors expressed herein do not
necessarily state or reflect those of the United States Government or any
agency thereof, or The Regents of the University of California.
\end{quotation}
\end{scriptsize}

\vskip 2in

\begin{center}
\begin{small}
{\it Lawrence Berkeley National Laboratory
is an equal opportunity employer.}
\end{small}
\end{center}

\newpage
\renewcommand{\thepage}{\arabic{page}}
\setcounter{page}{1}

\renewcommand{\arraystretch}{2}

\section{Introduction}
Studying the properties of the Higgs boson,
 once it is
discovered in future particle accelerators,
will be the prime tool to experimentally probe the
details of the electroweak symmetry breaking
mechanism
 in the Standard Model.
Of particular interest will be the Higgs boson decay
into bottom quarks, since the decay channel
$H\rightarrow b\bar{b}$ dominates in the intermediate
  Higgs mass range $M_H < 2 M_W$.
This process will be even more important, if
possible hints for new physics effects in the
 reported descrepancy \cite{Olc95}
 between the measured
partial Z boson width $R_b$ into bottom quarks
and its theoretical prediction
should happen to
substantiate.
Similar effects might then also be visible in Higgs
 decays $H \rightarrow b\bar{b}$
and emphasize the need
 for  precise Standard Model predictions
to $\Gamma(H\rightarrow b\bar{b})$.

 As a consequence much work has been spent on the calculation
of radiative corrections to Higgs processes in the past
 and excellent reviews on Higgs phenomenology
can be found in the literature
\cite{GunHabKanDaw90,Kni94a}.
Previous works concerning the partial rate
$\Gamma(H\rightarrow b\bar{b})$
 include electroweak one loop corrections
\cite{BarVilKhr91,Kni92,DabHol92},
 the calculations of universal
and nonuniversal corrections
 of the order
${\cal O}(\alpha_s G_F M_t^2)$
\cite{KniSir93,Kni94b,KwiSte94,KniSpi94},
and recently even a  three loop
${\cal O}(\alpha_s^2 G_F M_t^2)$
calculation was presented
\cite{KniSte95a,KniSte95b}.
Nonuniversal
corrections to the vertex $Hb\bar{b}$
involve the virtual
top quark through Higgs ghost exchange. Their
  top mass enhancement $\propto m_t^2$
 due to Yukawa couplings distinguishes them from
similar vertices of the Higgs boson to other
quark flavours.

In our earlier work \cite{KwiSte94}
 the
diagrams of Figure \ref{fig1}
were considered in the
 heavy top limit $M_t^2 \gg M_H^2$.
The two loop
${\cal O}(\alpha_s G_F M_t^2)$
 relation between the bare  mass $m_0$ and the
on-shell (OS) mass
$M_b$ of the bottom quark was presented
and the corrections to the partial Higgs width
were expressed in terms of $M_b$.

The $\msbar$ renormalization scheme  on the
other hand is the commonly used renormalization
prescription in higher order QCD calculations.
Apart from calculational convenience
its  concept of the running
 bottom mass $\bar{m}_b$ allows the
absorption of large
 logarithms $\ln (M_b^2/M_H^2)$
(see e.g. \cite{BraLev80,Sak80,KatKim94,Sur94a})
 and
causes the perturbation series to converge more
rapidly  than in the OS scheme.
It is therefore of obvious interest to
adopt the notion of the
running $\msbar$ mass $\bar{m}_b$
in the Higgs decay rate also for the
case when electroweak
 corrections are included.

 For this reason we have calculated the two loop relation
of order
${\cal O}(\alpha_s G_F M_t^2)$
between the on-shell mass and the $\msbar$ renormalized
mass of the bottom quark (for a discussion at the
one loop level see \cite{HemKni95}).
This transformation from one renormalization scheme
to the other
allows to express
$\Gamma(H\rightarrow b\bar{b})$ to the order
 ${\cal O}(\alpha_s G_F M_t^2)$  throughout
in terms of the running mass $\bar{m}_b$.

The problem is approached in four different ways.
All methods are leading to the same answer and thus
provide  powerful crosschecks beyond the
 standard consistency checks such as gauge invariance.

In order to introduce our notation let us start
from the bare Lagrangian
 and consider the bare
fermion propagator for the bottom quark
\beq \EQN{bareprop}
S_0^{-1} =
 i ( m_0-p\hspace{-0.45em}/- m_0\Sigma_S^0-p\hspace{-0.45em}/
    \Sigma_V^0 )
.\eeq
We have not written the term
$p\hspace{-0.45em}/\gamma_5 \Sigma_A^0$
for notational simplicity, since
for all quark mass relations
below
 $\Sigma_A$ becomes relevant only in higher
 order  electroweak corrections ${\cal O}(G_F^2)$,
 which we do not consider in this work.
We quote our previous result for the
${\cal O}(\alpha_s G_F M_t^2)$
bottom  pole mass
\beq \EQN{polemass}
M_b =
 m_0\frac{1-\Sigma_S^0}{1+\Sigma_V^0}
.\eeq
in Eq.(\ref{apppolemass}) of the appendix,
where for later convenience
 $\Sigma_S^0$
and $\Sigma_V^0$ are expressed in terms of the
$\msbar$ masses $\bar{m}_b$ and  $\bar{m}_t$.

By rescaling its parameters the bare Lagrangian
can be written as the sum of the renormalized Lagrangian
and the counterterm Lagrangian.
Our interest focuses on
the renormalization
 constants $Z_2$ and $Z_m$ relating the bare wavefunction and
mass of the bottom quark to their renormalized
equivalents
\beq\EQN{rescale}
\Psi_0 = Z_2^{1/2} \Psi,\;\;\; m_0 = Z_m \bar{m}_b
.\eeq
Here we adopt the $\msbar$ renormalization scheme as
is indicated through bars.
The renormalized bottom quark propagator accordingly reads
\beq \EQN{renprop}
\ba{rl}\dsp
S_R^{-1} =
& \dsp
Z_2 S_0^{-1}
\\ \dsp
=
& \dsp
 i \Big( \bar{m}_b-p\hspace{-0.45em}/
    - \bar{m}_b\bar{\Sigma}_S-p\hspace{-0.45em}/\bar{\Sigma}_V
    +(Z_2Z_m-1)\bar{m}_b
    -p\hspace{-0.45em}/ (Z_2-1)\Big)
\ea
\eeq

\noindent
For the determination of the $\msbar$ bottom mass we
perform our calculations  according to the following
 different strategies.

In Section 2.1
the overall counterterms
$\bar{\Sigma}_S^{CT}, \bar{\Sigma}_V^{CT}$ to the
bottom selfenergy
are computed  in the $\msbar$ scheme.
With
\beq\EQN{renkonst}
\ba{ll} \dsp
Z_2
& \dsp
= 1-\bar{\Sigma}_V^{CT}
\\ \dsp
Z_2Z_m
& \dsp
= 1+\bar{\Sigma}_S^{CT}
\ea
\eeq
 one obtains the relation between the $\msbar$ and bare masses
\beq\EQN{msm0}
\bar{m}_b = m_0\frac{1-\bar{\Sigma}_V^{CT}}{1+\bar{\Sigma}_S^{CT}}
.\eeq
In combination with Eq.(\ref{polemass}) this leads to
 the transformation rule between
OS- and $\msbar$ masses of the bottom quark.
\beq\EQN{mosms}
M_b = \bar{m}_b\frac{(1+\bar{\Sigma}_S^{CT})(1-\Sigma_S^0)}
                    {(1-\bar{\Sigma}_V^{CT})(1-\Sigma_V^0)}
.\eeq

In Section 2.2 a different approach is used to verify
the findings of Section 2.1.
The renormalized bottom quark propagator
Eq.(\ref{renprop})
is rewritten in the form
\beq\EQN{propfin}
S_R^{-1} =
i
\left\{
\bar{m}_b \left(1-\bar{\Sigma}_S+\bar{\Sigma}_S^{CT}
          \right)
-p\hspace{-0.45em}/
           \left(1+\bar{\Sigma}_V-\bar{\Sigma}_V^{CT}
           \right)
\right\}
\eeq
with $\bar{\Sigma}_S = Z_2Z_m\Sigma_S^0,\;
\bar{\Sigma}_V = Z_2\Sigma_V^0$.
We check by explicit calculation
of the finite parts of the bottom quarks
self  energies
$\bar{\Sigma}_S^{fin},\bar{\Sigma}_V^{fin}$
that the relation
\beq\EQN{mfin}
\ba{rl}\dsp
M_b =
& \dsp
\bar{m}_b\frac{(1-\bar{\Sigma}_S+\bar{\Sigma}_S^{CT})}
              {(1+\bar{\Sigma}_V-\bar{\Sigma}_V^{CT})}
\\ \dsp
=
& \dsp
\bar{m}_b\frac{(1-\bar{\Sigma}_S^{fin})}
              {(1+\bar{\Sigma}_V^{fin})}
\ea
\eeq
is indeed equivalent to the prescription
Eq.(\ref{mosms}).

In Section 2.3 our problem is considered from
a third point of view, which becomes transparent
by expressing the renormalized quark propagator
in the following form
\beq\EQN{propren}
S_R^{-1} =
 i Z_2 \Big(Z_m \bar{m}_b-p\hspace{-0.45em}/
    - Z_m\bar{m}_b\Sigma_S^0-p\hspace{-0.45em}/\Sigma_V^0 \Big)
.\eeq
This expression is finite if the bare parameters in
$\Sigma_{S,V}^0$ are substituted in favour of the
renormalized ones. One therefore can solve for
$Z_2$ and $Z_m$ recursively, i.e. loop by loop.
The renormalization constant $Z_m$ leads then to
the same result for $\bar{m}_b$ as in the
previous sections.

Finally we demonstrate in Section 2.4 that
another simple derivation of the result is possible,
based on the earlier determination of the bottom pole mass
and  leading to
the same $\bar{m}_b$ again.

The  results are then applied in Section 3 to the partial
decay rate $\Gamma(H\rightarrow b\bar{b})$. The
numerical size of the corrections are given and
the renormalization scheme dependence is discussed.

\section{Calculation of the $\msbar$ Renormalized
Bottom Mass}
\subsection{Approach 1: Counterterms}

We calculate the $\msbar$ counterterms on a
graph by graph basis in this section.
Per definition of the $\msbar$ scheme
counterterm vertices  consist
of pole terms only and are therefore  easier to
compute than full diagrams.
The integrals represented by the graphs in Figure \ref{fig1}
involve several mass scales. Via electroweak
interactions the top quark and the Higgs ghost come into
play with their respective scales $M_t$ and $M_W$.
In the heavy top limit $M_t^2\rightarrow \infty$
one has $M_W^2 \ll M_t^2$. Since we consider
only the leading term $\propto
M_t^2$ in the
power series of the inverse top mass, one can
neglect $M_W$ right from the beginning. As a consequence
the electroweak gauge parameter drops out trivially.
The heavy mass expansion
\cite{PivTka84,GorLar87,CheSmi87,Smi91}
has developed into a well established technique
and was sucessfully used in a number of
applications. For a more detailed description
 the reader is referred for example
to \cite{CheKueKwi95}.
 The main virtue of this method
 is the factorization of a multiloop integral
containing the heavy top quark into an integral
with less number of loops and
massive tadpole integrals.

This decomposition is operative in our problem as well.
Two loop integrals eventually factorize into a
one loop tadpole and a one loop propagator integral,
where the latter involves two scales, namely the
bottom mass and the external momentum.
However, being interested in the pole parts only,
the matter simplifies even more. Since the
pole terms are independent of masses and momenta,
one can conveniently nullify either of them.
Care must be taken that no spurious infrared
divergencies are introduced in this way.
In our case we have obtained the pole parts
to $\Sigma_S$ by setting the external momentum
to zero, thus reducing the massive propagator
integral to a tadpole integral. Similarly,
for the computation of $\Sigma_V$ the bottom
mass is nullified. The resulting massless
propagator integral is conveniently computed
with the help of MINCER
\cite{LarTkaVer91}
which is based
on the symbolic manipulation program FORM
\cite{Ver91}.

The counterterms of the one loop diagrams
``QCD'' and ``EW'' of Figure \ref{fig1} are simply
given by their pole terms obtained in the above
described manner.
On the two loop level the situation is somewhat more
involved, since the diagrams ``IN'', ``OUT''
and ``LEFT'' contain ultraviolet divergent subgraphs.
As is indicated in Figure \ref{fig2}, these subdivergences
have to be subtracted in order to arrive at the
overall divergence of the corresponding diagrams.
The removal of the subdivergences results in
local counterterm vertices, which we list in the
appendix. It can be seen that indeed all logarithms
have dropped out.

Whereas the counterterms are still gauge dependent,
the QCD gauge parameter $\xi_s$ cancels in
the following
expression for the bottom mass:
\beq\EQN{ctmsm0}
\ba{rl}\dsp
\bar{m}_b =
& \dsp
 m_0\frac{1-\bar{\Sigma}_V^{CT}}{1+\bar{\Sigma}_S^{CT}}
\\ \dsp
=
& \dsp
m_0 \left(
1+\api\frac{1}{\epsilon}
+ \bar{x}_t\frac{3}{2}\frac{1}{\epsilon}
+ \api\bar{x}_t\frac{2}{\epsilon}
    \right)
\ea
\eeq
with $\bar{x}_t = G_F \bar{m}_t^2/8\sqrt{2}\pi^2$.
This leads to the transformation between the
pole and the $\msbar$ mass of the bottom quark
\beq\EQN{ctmosms}
\ba{rll}\dsp
M_b =
& \dsp
\bar{m}_b
& \dsp
\frac{(1+\bar{\Sigma}_S^{CT})(1-\Sigma_S^0)}
              {(1-\bar{\Sigma}_V^{CT})(1-\Sigma_V^0)}
\\ \dsp
=
& \dsp
\bar{m}_b
& \dsp
\Bigg\{
1+\api\left(
\frac{4}{3}
+\ln\frac{\mu^2}{\bar{m}_b^2}
      \right)
+ \bar{x}_t
  \left(
\frac{5}{4}
+\frac{3}{2}\ln\frac{\mu^2}{\bar{m}_t^2}
  \right)
\\ \dsp
&
& \dsp
+ \api \bar{x}_t
 \Bigg(
\frac{9}{2}-4\zeta(2)
+\frac{5}{2}\ln\frac{\mu^2}{\bar{m}_t^2}
+\frac{5}{4}\ln\frac{\mu^2}{\bar{m}_b^2}
+\frac{3}{2}\ln^2\frac{\mu^2}{\bar{m}_t^2}
+\frac{3}{2}\ln\frac{\mu^2}{\bar{m}_t^2}
\ln\frac{\mu^2}{\bar{m}_b^2}
 \Bigg)
 \Bigg\}.
\ea
\eeq

\subsection{Approach 2: Finite parts}

As a cross check of the result Eq.(\ref{ctmosms})
we now want to recalculate it in a different way,
namely by employing only the finite parts of
the corresponding bottom self energy graph as
given in Eq.(\ref{mfin}).
The finite part of a diagram
\beq\EQN{fin}
\bar{\Sigma}_{S,V}^{fin} =
\bar{\Sigma}_{S,V}^{full}  -
\bar{\Sigma}_{S,V}^{sub} -
\bar{\Sigma}_{S,V}^{CT}
\eeq
is obtained
by subtracting the overall counterterm
$\bar{\Sigma}_{S,V}^{CT}$
and the counterterm
with the subdivergence
$\bar{\Sigma}_{S,V}^{sub}$
 from the full
diagram
$\bar{\Sigma}_{S,V}^{full}$.
Pictorially this procedure is
visualized in Figure \ref{fig3}.
Notice that
 $\bar{\Sigma}_{S,V}^{sub}$ contains
both pole and finite terms.
One therefore cannot use the nullification procedure
of the previous section to simplify the
calculation.

Instead it is possible to simplify
integrals  by evaluating them
on the mass shell $p^2=M_b^2$
\cite{GraBroGraSch90} using the   expansion
\beq\EQN{finexp}
\ba{rl}\dsp
\bar{\Sigma}^{full}_{S,V}\left(\frac{\bar{m}_b^2}{p^2}\right) =
& \dsp
\bar{\Sigma}^{full}_{S,V}(1) +
\left(\frac{\bar{m}_b^2}{p^2}-1\right)
\bar{\Sigma}_{S,V}^{\prime}(1)
\\ \dsp
=
& \dsp
\bar{\Sigma}^{full}_{S,V}(1) +
2 \left(\bar{\Sigma}_{S}+\bar{\Sigma}_{V}\right)
\bar{\Sigma}_{S,V}^{\prime}(1)
.\ea
\eeq
The derivatives $\bar{\Sigma}_{S,V}^{\prime}
\equiv \partial \bar{\Sigma}_{S,V}
/\partial(\bar{m}_b^2/p^2)$
may be conveniently obtained through derivations
with respect to $\bar{m}_b$, thus raising the power
in the denominator of the integrand.
This procedure may also be applied for the
calculation of subdivergence counterterms, where
the corresponding expansion reads
\beq\EQN{finsub}
 \bar{\Sigma}_{S,V}^{sub} \left(\frac{\bar{m}_b^2}{p^2}\right)
=\bar{\Sigma}_{S,V}^{sub}(1) +
2 \left(\bar{\Sigma}_{S}^{CT}+\bar{\Sigma}_{V}^{CT}\right)
\bar{\Sigma}_{S,V}^{\prime}(1)
.\eeq

The expressions for the finite parts of the various
 contributions are listed in the appendix.
They lead to the relation between pole and
$\msbar$ bottom mass
\beq\EQN{finmosms}
\ba{rll}\dsp
M_b =
& \dsp
\bar{m}_b
& \dsp
\frac{(1-\bar{\Sigma}_S^{fin})}
              {(1+\bar{\Sigma}_V^{fin})}
\\ \dsp
=
&\dsp
\bar{m}_b
& \dsp
\Bigg\{
1+\api\left(
\frac{4}{3}
+\ln\frac{\mu^2}{\bar{m}_b^2}
      \right)
+ \bar{x}_t
  \left(
\frac{5}{4}
+\frac{3}{2}\ln\frac{\mu^2}{\bar{m}_t^2}
  \right)
\\ \dsp
&
& \dsp
+ \api \bar{x}_t
 \Bigg(
\frac{9}{2}-4\zeta(2)
+\frac{5}{2}\ln\frac{\mu^2}{\bar{m}_t^2}
+\frac{5}{4}\ln\frac{\mu^2}{\bar{m}_b^2}
+\frac{3}{2}\ln^2\frac{\mu^2}{\bar{m}_t^2}
+\frac{3}{2}\ln\frac{\mu^2}{\bar{m}_t^2}
\ln\frac{\mu^2}{\bar{m}_b^2}
 \Bigg)
 \Bigg\}
\ea
\eeq
We find agreement with Eq.(\ref{ctmosms}).

\subsection{Approach 3: Renormalization Constants}
In our third method we proceed along a path which
deals directly with the renormalization constants
$Z_2$ and $Z_m$. To explain how
 both $Z_2,Z_m$ are computed iteratively
loop by loop, it is convenient to consider the
the renormalized fermion propagator in the
 following form:
\beq\EQN{rencstprop}
S_R^{-1} =
 i Z_2 \Big(Z_m \bar{m}_b-p\hspace{-0.45em}/
    - Z_m\bar{m}_b\Sigma_S^0-p\hspace{-0.45em}/
   \Sigma_V^0 \Big)
\eeq
Here the bare bottom selfenergies
$\Sigma_{S,V}^0=\Sigma_{S,V}^{(1)0}(m_{b,0},m_{t,0})
+\Sigma_{S,V}^{(2)0}(m_{b,0},m_{t,0}) $
receive contributions from the one and two loop
diagrams of Figure \ref{fig1}.
The explixcit  arguments shall emphasize that
all parameters are understood as bare quantities.
 In general the parameterlist
would also include coupling constants, gauge parameters
etc. If we now substitute the
bare masses in favour of
their renormalized counterparts at a given loop level,
the functional
form of the selfenergies does not change in that
given order, but additional contributions of higher
order are induced:
\beq\EQN{rencstind}
\Sigma_{S,V}^{(1)0}(m_{b,0},m_{t,0}) =
\Sigma_{S,V}^{(1)0}(\bar{m}_b,\bar{m}_t) +
\Sigma_{S,V,{\rm ind}}^{(2)}
\eeq
Let us first consider the one loop case.
 Having  expressed Eq.(\ref{rencstprop}) entirely in
terms of renormalized quantities, the renormalization
constants $Z_2,Z_m$ must be such that the inverse
quark propagator is finite and, stated more precisely for
 the $\msbar$ scheme, that the poles
cancel. According to the Lorentz structure this
results in two equations
\beq\EQN{rencsteq}
\ba{rll}\dsp
\bar{m}_b Z_2 Z_m\left(
1-   \Sigma_{S}^{(1)0}(\bar{m}_b,\bar{m}_t)
                \right)
 & \dsp
\stackrel{!}{=}
& \dsp
{\rm finite}
\\ \dsp
p\hspace{-0.45em}/ Z_2 \left(
1+  \Sigma_{V}^{(1)0}(\bar{m}_b,\bar{m}_t)
                \right)
 & \dsp
\stackrel{!}{=}
& \dsp
{\rm finite}
, \ea
\eeq
which can be solved for $Z_2$ and $Z_m$.
The solution for $Z_m$  leads to the
one loop result for the $\msbar$ mass $\bar{m}_b$.

The procedure can then be repeated for the two loop case.
Besides the two loop result $\Sigma_{S,V}^{(2)0}$ also
the  induced second order terms
$\Sigma_{S,V,{\rm ind}}^{(2)}$
 from the transition to the
renormalized parameters at the one loop iteration
have to be taken into account. Solving the corresponding
system of equations gives the renormalization constants
at the two loop level. Inversion of Eq.(\ref{rescale})
with
\begin{eqnarray}
Z_m &=& 1 - \bar{x}_t\frac{3}{2\epsilon}
          - \frac{\alpha_s}{\pi}\frac{1}{\epsilon}
          + \frac{\alpha_s}{\pi}\bar{x}_t\left(
                \frac{3}{\epsilon^2}
               -\frac{2}{\epsilon}         \right).
\end{eqnarray}
indeed confirms the result
Eq.(\ref{ctmsm0}).

\subsection{Approach 4: Derivation from OS Mass}

Having approached the problem from three different sides,
let us demonstrate, how the $\msbar$ bottom mass
can be  derived in another elegant manner.
We start with the following ansatz
 for the relation between
the bare mass and the $\msbar$ mass of the
bottom quark
(a similar method was used in
\cite{NacWet81})
and insert it into Eq.(\ref{polemass}):
\beq\EQN{deransatz}
M_b =
\bar{m}_b
\left\{
1+ \api\frac{a}{\epsilon}
+ x_{t,0}\frac{b}{\epsilon}
+ \api x_{t,0}
  \left(
        \frac{c}{\epsilon^2}+
        \frac{d}{\epsilon}
  \right)
\right\}
\frac{1-\Sigma_S^0}{1+\Sigma_V^0}
\eeq
The bare top mass
in $x_{t,0}$ is substituted
 through the renormalized $\msbar$ mass
and the terms following the curly bracket
are taken  from
the pole mass calculation
in Eq.(\ref{apppolemass}).

The crucial step is to require that the pole mass
$M_b$ on the LHS as a physical quantity must
be finite. This translates into the
requirement that all coefficients of
$1/\epsilon$ poles on the RHS must vanish.
Thus one obtains
 four equations which can be solved for
the unknown coefficients $a,b,c,d$.
 Two additional equalities follow from the fact
that the logarithms of the pole terms cancel
separately and serve as a consistency check for
the solutions
\beq\EQN{dersolutions}
a=-1,\;\; b=-\frac{3}{2},\;\;
c=0,\;\; d=-2
.\eeq
Insertion into Eq.(\ref{deransatz})
produces again the result
Eq.(\ref{ctmosms}) and Eq.(\ref{finmosms}).

\section{Application to the Higgs Decay $H\rightarrow b\bar{b}$}
In this section we apply our result  to the partial Higgs boson
decay rate
\cite{KwiSte94,KniSpi94}
\beq\EQN{appgos}
\Gamma(H\rightarrow b\bar{b}) =
\Gamma_0 M_b^2 \left\{
1+X_t+\api X_t\left(
-1-4\zeta(2)-2\ln\frac{M_H^2}{M_b^2}
             \right)
                \right\}
\eeq
where $\Gamma_0=3G_FM_H/4\sqrt{2}\pi$,
$X_t=G_F M_t^2/8\sqrt{2}\pi^2$
and the
renormalization scale is chosen as $\mu^2=M_H^2$.
For our following numerical discussion we use
as input values
a bottom pole mass of $M_b=4.7$ GeV and a
top mass of $M_t=176$ GeV. Based on
$\Lambda^{(5)}_{QCD}=233$ MeV the running strong
coupling constant ranges between
$\alpha_s(M_H=70\mbox{ GeV})=0.125$ and
$\alpha_s(M_H=130\mbox{ GeV})=0.114$ where
$\alpha_s(\mu)$ is defined for five active flavours.
We now express  the above formula for the
width  in terms of $\msbar$ masses and obtain
\beq\EQN{appgms}
\ba{rl}\dsp
\Gamma(H\rightarrow b\bar{b}) =
\Gamma_0 \bar{m}_b^2
& \dsp
\Bigg\{ 1 + \bar{x}_t\left(
 \frac{7}{2}+3\ln\frac{M_H^2}{\bar{m}_t^2}
                     \right)
\\ \dsp
& \dsp
+\api\bar{x}_t\left(
\frac{175}{6}-12\zeta(2)
+26\ln\frac{M_H^2}{\bar{m}_t^2}
+3\ln^2\frac{M_H^2}{\bar{m}_t^2}
               \right)
\Bigg\}
.\ea
\eeq
Notice that  the transformation of
Eq.(\ref{appgos}) implies that
 the first order QCD corrections
$\Gamma_0 M_b^2(\alpha_s/\pi)[3-2\ln(M_H^2/M_b^2)]$
give rise to a contribution of order
${\cal O}(\alpha_s G_F \bar{m}_t^2)$ as well.
Based on the given values for the pole masses
the corresponding running masses amount to
$\bar{m}_b=2.84/2.69$ GeV and
$\bar{m}_t=179.44/170.04$ GeV for
$M_H=70/130$ GeV.
In Figure \ref{fig4} we plot the corrections of
orders $G_Fm_t^2$ and
$G_Fm_t^2+\alpha_sG_Fm_t^2$ according to
Eqs.(\ref{appgos}) and (\ref{appgms}).
The curves  are strongly
characterized by the linear rise in $M_H$
due to  the overall factor.
For the on-shell result the QCD screening
of the leading electroweak corrections is
clearly visible. The $\msbar$ curves indicate
that the two loop contribution is less
important than for the OS scheme and
suggest a better convergence
behaviour of the
perturbation series.

An inspection of Eq.(\ref{appgms}) reveals
that all large
logarithms $\ln(M_H^2/M_b^2)$ have dropped
out. Their absorption into the running bottom
mass favours the use of the $\msbar$ mass
over the OS-mass.
There is no such strong preferance with respect
to the top mass, considering that the scales of the
Higgs and the top are not as far apart as
the Higgs and the bottom scales. Corresponding
logarithms $\ln(M_H^2/M_t^2)$ therefore cannot
be considered as particularly dangerous.
Instead one might tend to  use the
top pole mass as a quantity which is by definition
more feasible in experiments.
In this case the Higgs decay rate can be
rewritten into the following form
\beq\EQN{appmsos}
\ba{rl}\dsp
\Gamma(H\rightarrow b\bar{b}) =
\Gamma_0 \bar{m}_b^2
& \dsp
\Bigg\{ 1 + \frac{17}{3}\api
+X_t\left(
 \frac{7}{2}+3\ln\frac{M_H^2}{M_t^2}
                     \right)
\\ \dsp
& \dsp
+\api X_t\left(
\frac{167}{6}-12\zeta(2)
+17\ln\frac{M_H^2}{M_t^2}
-3\ln^2\frac{M_H^2}{M_t^2}
               \right)
\\ \dsp
& \dsp
+\left(\api\right)^2 \left(
30.717
-\frac{2}{3}\ln\frac{M_H^2}{M_t^2}
+\frac{1}{9}\ln^2\frac{M_H^2}{\bar{m}_b^2}
               \right)
\\ \dsp
& \dsp
+{\cal O}(\frac{\bar{m}_b^2}{M_H^2})
+{\cal O}(\alpha_s^2\frac{M_H^2}{M_t^2})
\Bigg\}
.\ea
\eeq
Several groups have contributed to the
calculation of QCD corrections which we have
included in the formula in first
\cite{BraLev80,Sak80,InaKub81,DreHik90,SurTka90}
 and massless second
order \cite{GorKatLar84,GorKatLarSur90}.
Quadratic bottom mass corrections in second order
\cite{Sur94a,CheKwi95} and top quark
contributions
\cite{Kni95,CheKwi95,LarRitVer95}
 are also available. We have not
written  these pieces into
Eq.(\ref{appmsos}), but included
them in our numerical analysis.

 As was noticed in \cite{CheKwi95}
the logarithms
$\alpha_s^2\ln^2(M_H^2/\bar{m}_b^2)$ originate from
flavour singlet type diagrams. They are not present
in the rate for the decay into hadrons, but are
introduced when the pure gluonic channel is subtracted.
In Figure \ref{fig5} the contributions coming from the orders
$\alpha_s, \alpha_s^2, G_FM_t^2, \alpha_sG_FM_t^2$
are compared, normalized to the Born term
$\Gamma_{{\rm Born}}=\Gamma_0\bar{m}_b^2$.
The electroweak corrections may carry different sign
depending on the Higgs mass. However, as compared
to the QCD corrections, where the first order
contributes about 20\% and the
second order about 5\% to the corrections,
the electroweak contributions are small.
With ${\cal O}(G_F M_t^2)=-6.6/5.4$ per mille
and  ${\cal O}(\alpha_s G_F M_t^2)=-4.3/-3.9$ per mille
for $M_H=70/130$ GeV these effects become relevant
for high precision experiments at the percent level.

\vspace{5ex}
{\bf Acknowledgments}

\noindent
 We would like to
thank K.G.Chetyrkin and J.H.K\"uhn
for helpful discussions.
A.K. thanks the
Deutsche Forschungsgemeinschaft for financial support
(grant Kw 8/1-1 ). Partial support by US DOE under
Contract DE-AC03-76SF00098 is gratefully acknowledged.

\appendix
\section{Appendix}
The result from \cite{KwiSte94} for the
relation between the pole mass and the
bare mass of the bottom quark
according to Eq.(\ref{polemass})
is reproduced below.
For convenient use in Section 2 the bare masses
in $\Sigma_S^0$ and $\Sigma_V^0$ are already
transformed into the $\msbar$ renormalized ones.
\beq\EQN{apppolemass}
\ba{rl}\dsp
M_b =
& \dsp
m_0\Bigg\{
1+\api\left(
\frac{1}{\epsilon}+\frac{4}{3}
+\ln\frac{\mu^2}{\bar{m}_b^2}
+\epsilon \left[
\frac{8}{3}+\frac{1}{2}\zeta(2)
+\frac{4}{3}\ln\frac{\mu^2}{\bar{m}_b^2}
+\frac{1}{2}\ln^2\frac{\mu^2}{\bar{m}_b^2}
          \right]
      \right)
\\ \dsp
& \dsp
+ \bar{x}_t
  \left(
\frac{3}{2}\frac{1}{\epsilon}+\frac{5}{4}
+\frac{3}{2}\ln\frac{\mu^2}{\bar{m}_t^2}
+\epsilon \left[
\frac{9}{8}+\frac{3}{4}\zeta(2)
+\frac{5}{4}\ln\frac{\mu^2}{\bar{m}_t^2}
+\frac{3}{4}\ln^2\frac{\mu^2}{\bar{m}_t^2}
          \right]
  \right)
\\ \dsp
& \dsp
+ \api \bar{x}_t
 \Bigg(
\frac{1}{\epsilon}
\left[
\frac{21}{4}
+\frac{3}{2}\ln\frac{\mu^2}{\bar{m}_t^2}
+\frac{3}{2}\ln\frac{\mu^2}{\bar{m}_b^2}
\right]
\\  & \dsp
\hphantom{+ \api \bar{x}_t(}
+\frac{77}{8}-\frac{5}{2}\zeta(2)
+\frac{15}{4}\ln\frac{\mu^2}{\bar{m}_t^2}
+\frac{13}{4}\ln\frac{\mu^2}{\bar{m}_b^2}
\\   & \dsp
\hphantom{+ \api \bar{x}_t(}
+\frac{9}{4}\ln^2\frac{\mu^2}{\bar{m}_t^2}
+\frac{3}{4}\ln^2\frac{\mu^2}{\bar{m}_b^2}
+\frac{3}{2}\ln\frac{\mu^2}{\bar{m}_t^2}
\ln\frac{\mu^2}{\bar{m}_b^2}
 \Bigg)
 \Bigg\}
\ea
\eeq

The overall counterterms are given for the
different diagrams:
\beq
\ba{rl} \dsp
\bar{\Sigma}_S^{CT}(QCD) =
& \dsp
\api\frac{1}{\epsilon}
\left(
 -1-\frac{1}{3}\xi_s
\right)
\\ \dsp
\bar{\Sigma}_V^{CT}(QCD) =
& \dsp
\api\frac{1}{\epsilon}
 \frac{1}{3}\xi_s
\\ \dsp
\bar{\Sigma}_S^{CT}(EW) =
& \dsp
\bar{x}_t\frac{-2}{\epsilon}
\\ \dsp
\bar{\Sigma}_V^{CT}(EW) =
& \dsp
\bar{x}_t\frac{1}{2\epsilon}
\ea
\eeq

\beq
\ba{rl} \dsp
\bar{\Sigma}_S^{CT}({\rm IN}) =
& \dsp
\api\bar{x}_t
\left\{
\frac{1}{\epsilon^2}
\left(
 1-\frac{1}{3}\xi_s
\right)
+\frac{1}{\epsilon}
\left(
 -\frac{1}{3}-\frac{1}{3}\xi_s
\right)
\right\}
\\ \dsp
\bar{\Sigma}_V^{CT}({\rm IN}) =
& \dsp
\api\bar{x}_t
\left\{
\frac{1}{\epsilon^2}
 \frac{1}{12}\xi_s
+\frac{1}{\epsilon}
 \frac{1}{24}\xi_s
\right\}
\ea
\eeq

\beq
\ba{rl} \dsp
\bar{\Sigma}_S^{CT}({\rm OUT}) =
& \dsp
\api\bar{x}_t
\left\{
\frac{1}{\epsilon^2}
\left(
 \frac{1}{2}+\frac{1}{6}\xi_s
\right)
+\frac{1}{\epsilon}
\left(
 -\frac{5}{6}-\frac{1}{6}\xi_s
\right)
\right\}
\\ \dsp
\bar{\Sigma}_V^{CT}({\rm OUT}) =
& \dsp
\api\bar{x}_t
\left\{
\frac{1}{\epsilon^2}
 \frac{1}{12}\xi_s
+\frac{1}{\epsilon}
\left(
-\frac{1}{8}\xi_s
\right)
\right\}
\ea
\eeq

\beq
\ba{rl} \dsp
\bar{\Sigma}_S^{CT}({\rm LEFT}) =
& \dsp
\api\bar{x}_t
\left\{
\frac{1}{\epsilon^2}
\left(
 \frac{5}{2}+\frac{5}{6}\xi_s
\right)
+\frac{1}{\epsilon}
\left(
 -\frac{1}{2}+\frac{1}{2}\xi_s
\right)
\right\}
\\ \dsp
\bar{\Sigma}_V^{CT}({\rm LEFT}) =
& \dsp
\api\bar{x}_t
\left\{
\frac{1}{\epsilon^2}
\left(
 -\frac{1}{2}-\frac{1}{3}\xi_s
\right)
+\frac{1}{\epsilon}
\left(
-\frac{1}{3}+\frac{1}{12}\xi_s
\right)
\right\}
\ea
\eeq

The finite parts of the one and two loop
contributions read as follows:
\beq
\ba{rl}\dsp
\bar{\Sigma}_S^{fin}({\rm 1-loop}) =
& \dsp
\api\left\{
-\frac{4}{3}-\frac{2}{3}\xi_s
+\left(
-1-\frac{1}{3}\xi_s
\right)
\ln\frac{\mu^2}{\bar{m}_b^2}
    \right\}
\\ \dsp
& \dsp
+\bar{x}_t\left\{
-2-2\ln\frac{\mu^2}{\bar{m}_t^2}
    \right\}
\ea
\eeq

\beq
\ba{rl}\dsp
\bar{\Sigma}_V^{fin}({\rm 1-loop}) =
& \dsp
\api\left\{
\frac{2}{3}\xi_s
+\frac{1}{3}\xi_s\ln\frac{\mu^2}{\bar{m}_b^2}
    \right\}
\\ \dsp
& \dsp
+\bar{x}_t\left\{
\frac{3}{4}+\frac{1}{2}\ln\frac{\mu^2}{\bar{m}_t^2}
    \right\}
\ea
\eeq

\beq
\ba{rl}\dsp
\bar{\Sigma}_S^{fin}({\rm 2-loop}) =
\api\bar{x}_t
& \dsp
\Bigg\{
-\frac{37}{6}-\frac{1}{2}\xi_s+4\zeta(2)
\\
& \dsp
+\left(
-3-\frac{1}{3}\xi_s
 \right)
\ln\frac{\mu^2}{\bar{m}_t^2}
+\left(
-2-\frac{2}{3}\xi_s
 \right)
\ln\frac{\mu^2}{\bar{m}_b^2}
\\
& \dsp
-2\ln^2\frac{\mu^2}{\bar{m}_t^2}
+\left(
-2-\frac{2}{3}\xi_s
 \right)
\ln\frac{\mu^2}{\bar{m}_t^2}
\ln\frac{\mu^2}{\bar{m}_b^2}
    \Bigg\}
\ea
\eeq

\beq
\ba{rl}\dsp
\bar{\Sigma}_V^{fin}({\rm 2-loop}) =
\api\bar{x}_t
& \dsp
\Bigg\{
\frac{2}{3}-\frac{1}{3}\xi_s
+\frac{1}{4}\xi_s
\ln\frac{\mu^2}{\bar{m}_b^2}
\\
& \dsp
+\left(
-\frac{1}{6}-\frac{2}{3}\xi_s
 \right)
\ln\frac{\mu^2}{\bar{m}_t^2}
\\
& \dsp
+\frac{1}{2}\ln^2\frac{\mu^2}{\bar{m}_t^2}
+\frac{1}{6}\xi_s
\ln\frac{\mu^2}{\bar{m}_t^2}
\ln\frac{\mu^2}{\bar{m}_b^2}
    \Bigg\}
\ea
\eeq

\begin{figure}[b]

\centerline{\bf Figure Captions}

\caption{ \label{fig1}
          Order ${\cal O}(\alpha_sG_FM_t^2)$ self
          energy diagrams for the bottom quark.
          Thin line: bottom quark, thick line: top quark,
          curly line: gluon, dashed line: Higgs ghost.
                          }

\caption{\label{fig2}
         Counterterm diagrams up to order ${\cal O}(\alpha_sG_FM_t^2)$,
          \hspace{5.5cm}
         }

\caption{\label{fig3}
         Finite terms for diagrams up to
         order ${\cal O}(\alpha_sG_FM_t^2)$.
          \hspace{5cm}
         }

\caption{\label{fig4}
   Corrections to $\Gamma(H\rightarrow b\bar{b})$
   in terms of pole masses (upper curves) and
   $\msbar$ masses (lower curves). The solid lines
   are the ${\cal O}(G_F m_t^2)$ contributions
   and the dashed lines
   the sum of  ${\cal O}( G_F m_t^2)$ and
   ${\cal O}(\alpha_s G_F m_t^2)$.
         }

\caption{\label{fig5}
  Corrections  $\Gamma(H\rightarrow b\bar{b})$ separately
  for the orders ${\cal O}(\alpha_s)$ (solid curve),
  ${\cal O}(\alpha_s^2)$ (dashed-dotted curve),
   ${\cal O}(G_F M_t^2)$ (dashed curve) and
   ${\cal O}(\alpha_s G_F M_t^2)$ (dotted curve).
         }

\end{figure}

\setcounter{figure}{0}

\newpage

%
%
\renewcommand{\arraystretch}{2.0}
\renewcommand{\baselinestretch}{2.0}
{\bf
%
%
\begin{figure}
 \begin{center}
  \begin{tabular}{ccc}
   \epsfxsize=3cm
   \leavevmode
   $\parbox{2cm}{\epsffile[189 311 423 442]{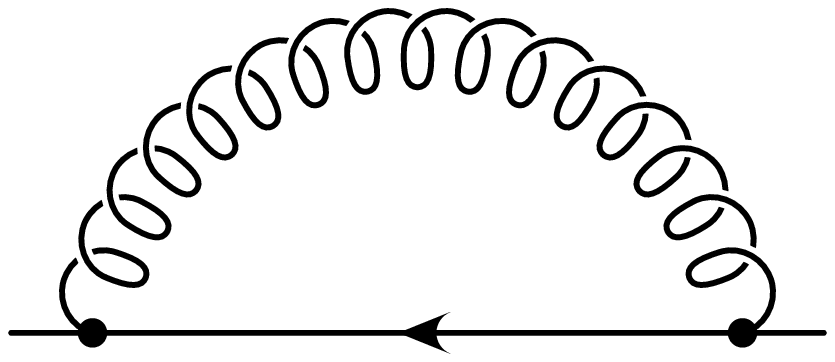}} \atop \mbox{QCD}$
   \epsfxsize=3cm
   \leavevmode
   $\parbox{2cm}{\epsffile[189 311 423 442]{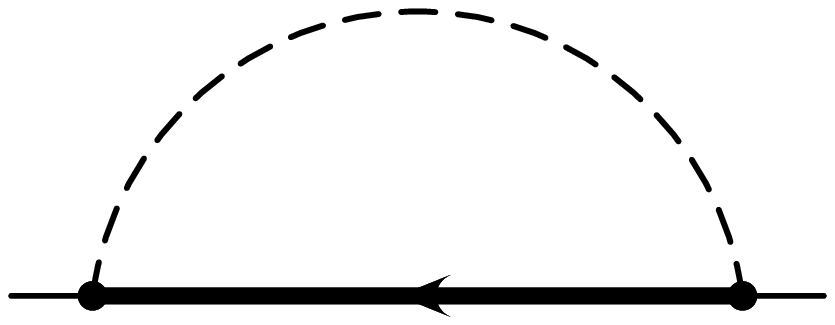}} \atop \mbox{EW}$
  \end{tabular}

  \begin{tabular}{ccc}
   \epsfxsize=3cm
   \leavevmode
   $\parbox{2cm}{\epsffile[189 299 423 434]{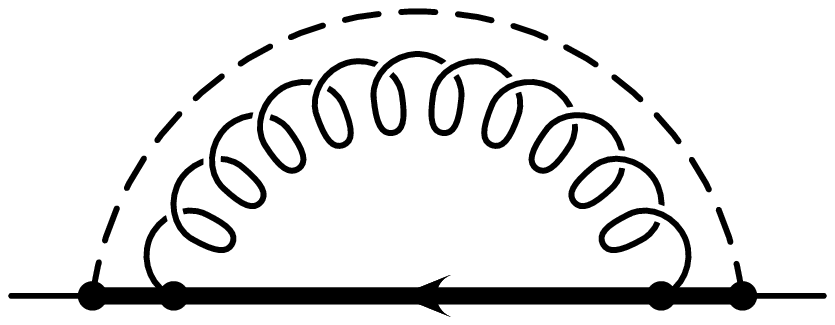}} \atop \mbox{IN}$
   &
   \epsfxsize=3cm
   \leavevmode
   $\parbox{2cm}{\epsffile[189 299 423 434]{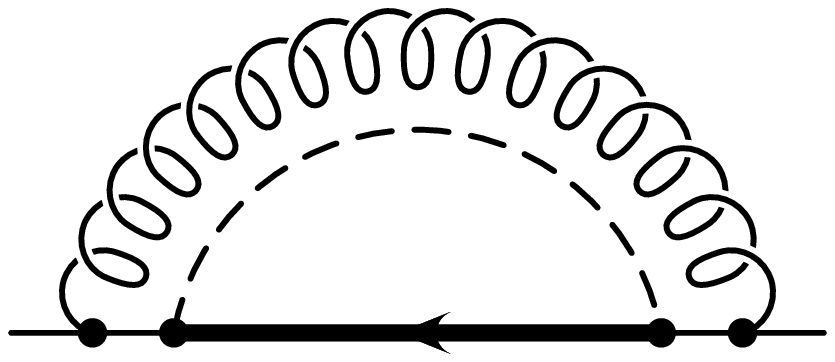}} \atop \mbox{OUT}$
   &
   \epsfxsize=3cm
   \leavevmode
   $\parbox{2cm}{\epsffile[189 299 423 434]{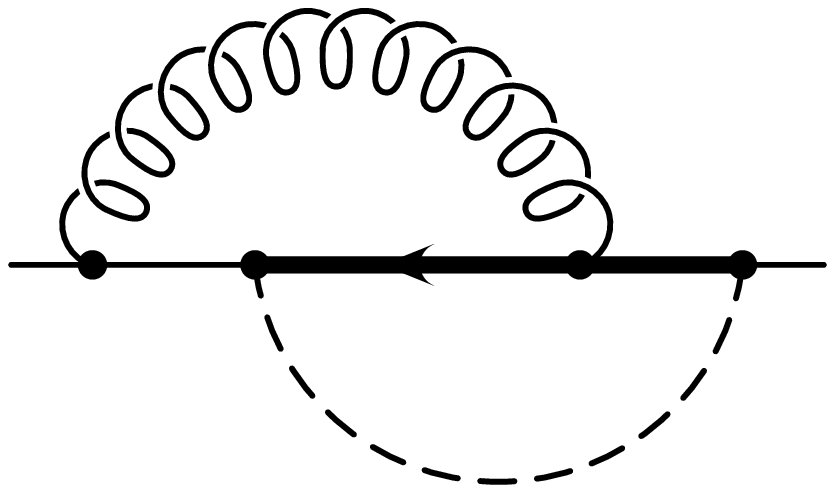}} \atop \mbox{LEFT}$
  \end{tabular}

\vspace{3ex}
Figure \ref{fig1}
 \end{center}
\end{figure}

%
%
\begin{figure}
 \begin{center}
  \begin{tabular}{ccc}
   \epsfxsize=3cm
   \leavevmode
   \parbox{2cm}{\epsffile[189 384 423 408]{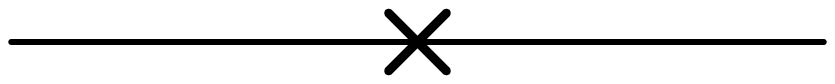}}
   &
   =\raisebox{3ex}{
     \begin{minipage}{3cm}
      {\Huge
       \begin{eqnarray}
        \sum_{\mbox{\scriptsize i$=$QCD, EW, IN} \atop
              \mbox{\scriptsize OUT, LEFT}}
        \nonumber
       \end{eqnarray}
      }
     \end{minipage}
    }
   &
   \epsfxsize=3cm
   \leavevmode
   $\parbox{2cm}{\epsffile[189 390 423 408]{mbct.ps}} \atop \mbox{i}$
  \end{tabular}

  \begin{tabular}{ccccc}
  &=&
  \parbox{4.5cm}{
   \epsfysize=2.5cm
   \leavevmode
   \epsffile[270 325 330 467]{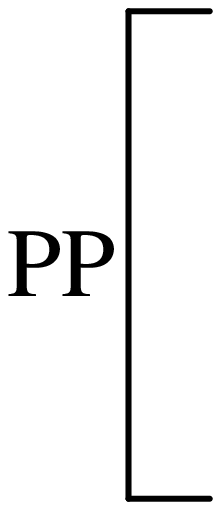}
   \epsfxsize=3cm
   \leavevmode
   \epsffile[189 311 423 442]{mbqcd.ps}
   \epsfysize=2.5cm
   \leavevmode
   \epsffile[282 325 306 467]{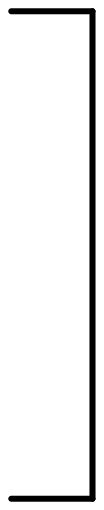}
  }
  &+&
  \parbox{4.5cm}{
   \epsfysize=2.5cm
   \leavevmode
   \epsffile[270 325 330 467]{mbPPobr.ps}
   \epsfxsize=3cm
   \leavevmode
   \epsffile[189 311 423 442]{mbew.ps}
   \epsfysize=2.5cm
   \leavevmode
   \epsffile[282 325 306 467]{mbcbr.ps}
  }
\\
  &+&
  \parbox{4.5cm}{
   \epsfysize=2.5cm
   \leavevmode
   \epsffile[270 325 330 467]{mbPPobr.ps}
   \epsfxsize=3cm
   \leavevmode
   \epsffile[189 311 423 442]{mbin.ps}
  }
   &--&
  \parbox{4.5cm}{
   \epsfxsize=3cm
   \leavevmode
   \epsffile[189 311 423 442]{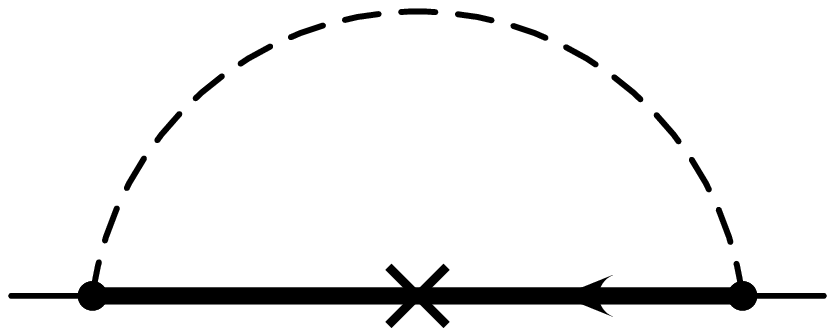}
   \epsfysize=2.5cm
   \leavevmode
   \epsffile[282 325 306 467]{mbcbr.ps}
  }
\\
  &+&
  \parbox{4.5cm}{
   \epsfysize=2.5cm
   \leavevmode
   \epsffile[270 325 330 467]{mbPPobr.ps}
   \epsfxsize=3cm
   \leavevmode
   \epsffile[189 311 423 442]{mbout.ps}
  }
   &--&
  \parbox{4.5cm}{
   \epsfxsize=3cm
   \leavevmode
   \epsffile[189 311 423 442]{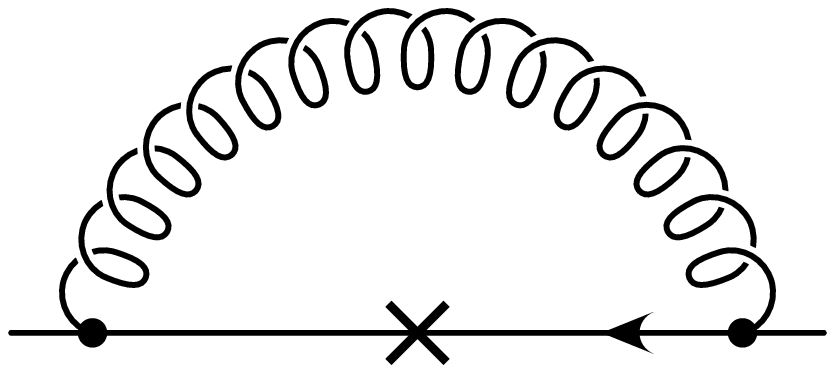}
   \epsfysize=2.5cm
   \leavevmode
   \epsffile[282 325 306 467]{mbcbr.ps}
  }
\\
  &+&
  \parbox{4.5cm}{
   \epsfysize=2.5cm
   \leavevmode
   \epsffile[270 325 330 467]{mbPPobr.ps}
   \epsfxsize=3cm
   \leavevmode
   \epsffile[189 311 423 442]{mbleft.ps}
  }
   &--&
  \parbox{4.5cm}{
   \epsfxsize=3cm
   \leavevmode
   \epsffile[189 349 423 442]{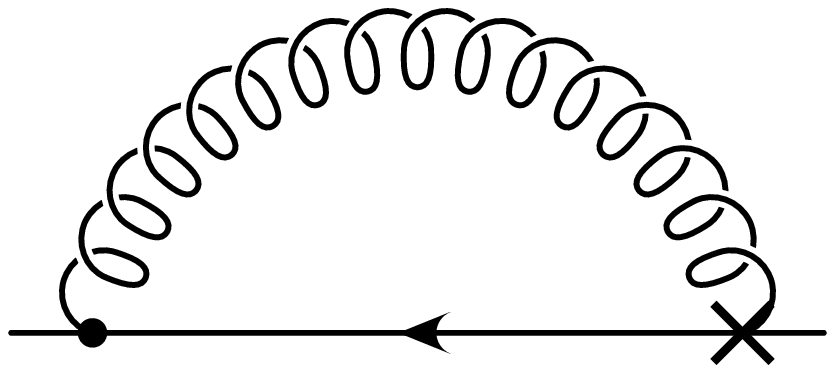}
  }
\\
  &&& -- &
  \parbox{4.5cm}{
   \epsfxsize=3cm
   \leavevmode
   \epsffile[189 311 423 442]{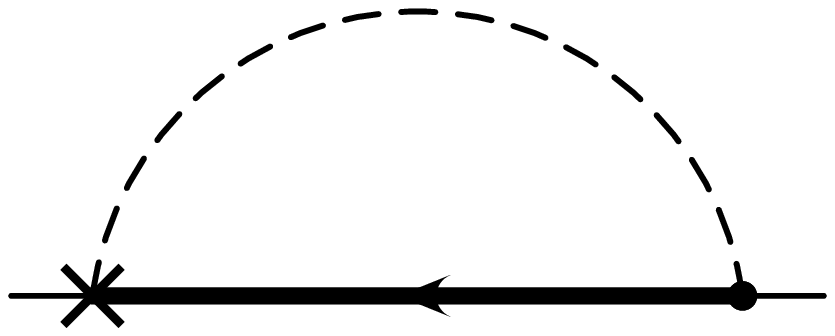}
   \epsfysize=2.5cm
   \leavevmode
   \epsffile[282 325 306 467]{mbcbr.ps}
  }
  \end{tabular}

\vspace{3ex}
Figure \ref{fig2}
 \end{center}
\end{figure}

%
%
\begin{figure}
 \begin{center}
  \begin{tabular}{ccc}
   \epsfxsize=3cm
   \leavevmode
   \parbox{2cm}{\epsffile[189 384 423 408]{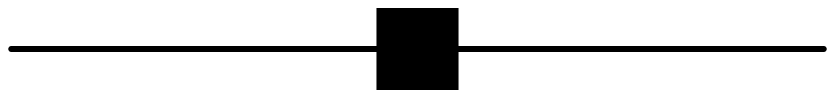}}
   &
   =\raisebox{3ex}{
     \begin{minipage}{3cm}
      {\Huge
       \begin{eqnarray}
        \sum_{\mbox{\scriptsize i$=$QCD, EW, IN} \atop
              \mbox{\scriptsize OUT, LEFT}}
        \nonumber
       \end{eqnarray}
      }
     \end{minipage}
    }
   &
   \epsfxsize=3cm
   \leavevmode
   $\parbox{2cm}{\epsffile[189 384 423 408]{mbfin.ps}} \atop \mbox{i}$
  \end{tabular}

  \begin{tabular}{ccccccc}
  &=&
  \parbox{3.2cm}{
   \epsfxsize=3cm
   \leavevmode
   \epsffile[189 311 423 442]{mbqcd.ps}
  }
  &--&
  \parbox{3.2cm}{
   \epsfxsize=3cm
   \leavevmode
   $\parbox{2cm}{\epsffile[189 384 423 408]{mbct.ps}} \atop \mbox{QCD}$
  }
\\
  &+&
  \parbox{3.2cm}{
   \epsfxsize=3cm
   \leavevmode
   \epsffile[189 311 423 442]{mbew.ps}
  }
  &--&
  \parbox{3.2cm}{
   \epsfxsize=3cm
   \leavevmode
   $\parbox{2cm}{\epsffile[189 384 423 408]{mbct.ps}} \atop \mbox{EW}$
  }
\\
  &+&
  \parbox{3.2cm}{
   \epsfxsize=3cm
   \leavevmode
   \epsffile[189 311 423 442]{mbin.ps}
  }
   &--&
  \parbox{3.2cm}{
   \epsfxsize=3cm
   \leavevmode
   \epsffile[189 311 423 442]{mbewct.ps}
  }
   &--&
  \parbox{3.2cm}{
   \epsfxsize=3cm
   \leavevmode
   $\parbox{2cm}{\epsffile[189 384 423 408]{mbct.ps}} \atop \mbox{IN}$
  }
\\
  &+&
  \parbox{3.2cm}{
   \epsfxsize=3cm
   \leavevmode
   \epsffile[189 311 423 442]{mbout.ps}
  }
   &--&
  \parbox{3.2cm}{
   \epsfxsize=3cm
   \leavevmode
   \epsffile[189 311 423 442]{mbqcdct.ps}
  }
   &--&
  \parbox{3.2cm}{
   \epsfxsize=3cm
   \leavevmode
   $\parbox{2cm}{\epsffile[189 384 423 408]{mbct.ps}} \atop \mbox{OUT}$
  }
\\
  &+&
  \parbox{3.2cm}{
   \epsfxsize=3cm
   \leavevmode
   \epsffile[189 311 423 442]{mbleft.ps}
  }
   &--&
  \parbox{3.2cm}{
   \epsfxsize=3cm
   \leavevmode
   \epsffile[189 349 423 442]{mbqcdct2.ps}
  }
\\
  &&& -- &
  \parbox{3.2cm}{
   \epsfxsize=3cm
   \leavevmode
   \epsffile[189 311 423 442]{mbewct2.ps}
  }
   &--&
  \parbox{3.2cm}{
   \epsfxsize=3cm
   \leavevmode
   $\parbox{2cm}{\epsffile[189 384 423 408]{mbct.ps}} \atop \mbox{LEFT}$
  }
  \end{tabular}

\vspace{3ex}
Figure \ref{fig3}
 \end{center}
\end{figure}

} 


\newpage

\begin{figure}
 \begin{center}
 \begin{tabular}{c}
   \epsfxsize=12.0cm
   \epsffile[50 280 520 600]{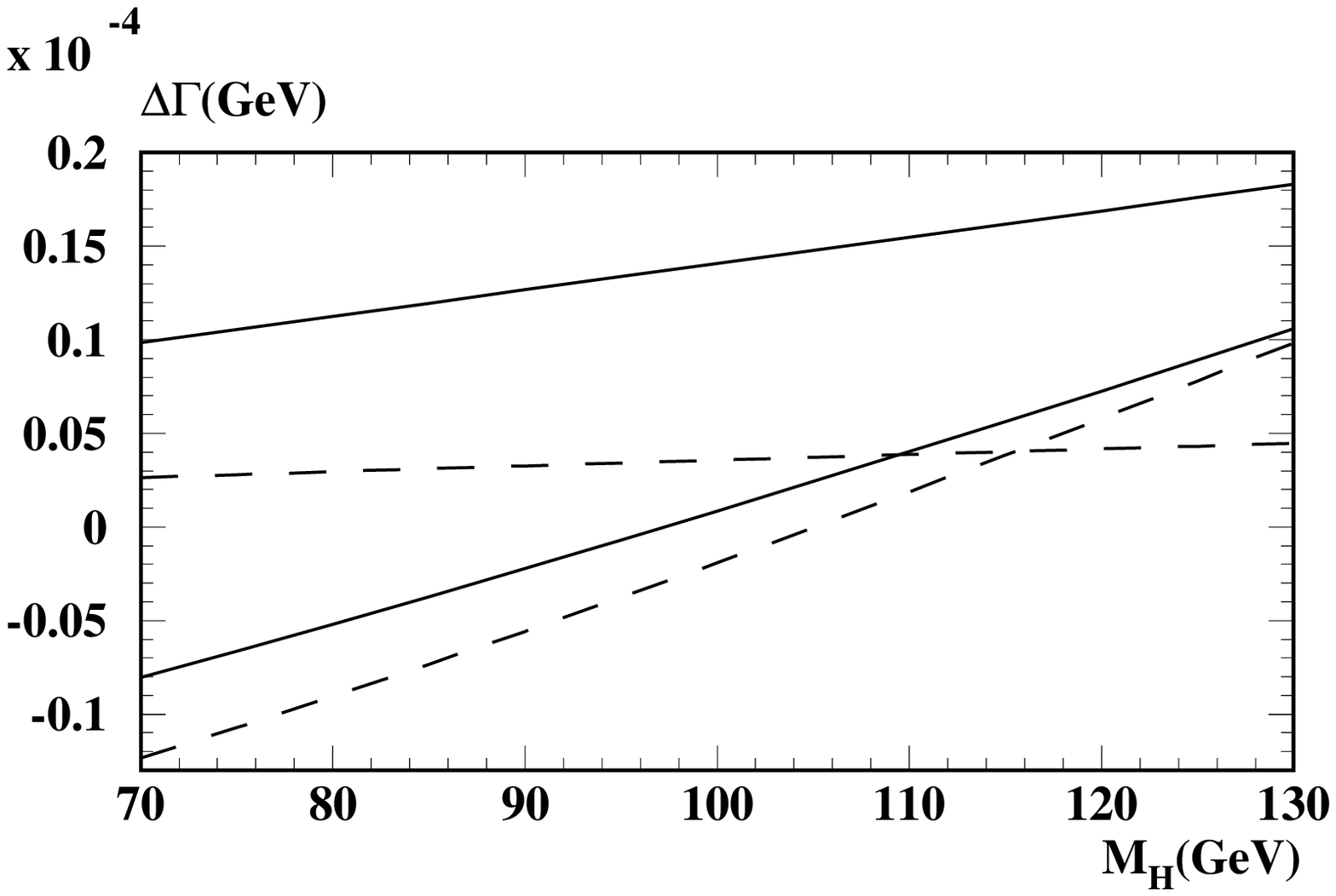}
 \end{tabular}

\vspace{3ex}
{\bf Figure \ref{fig4}}
 \end{center}
\end{figure}

\newpage

\begin{figure}
 \begin{center}
 \begin{tabular}{c}
   \epsfxsize=12.0cm
   \epsffile[50 280 520 600]{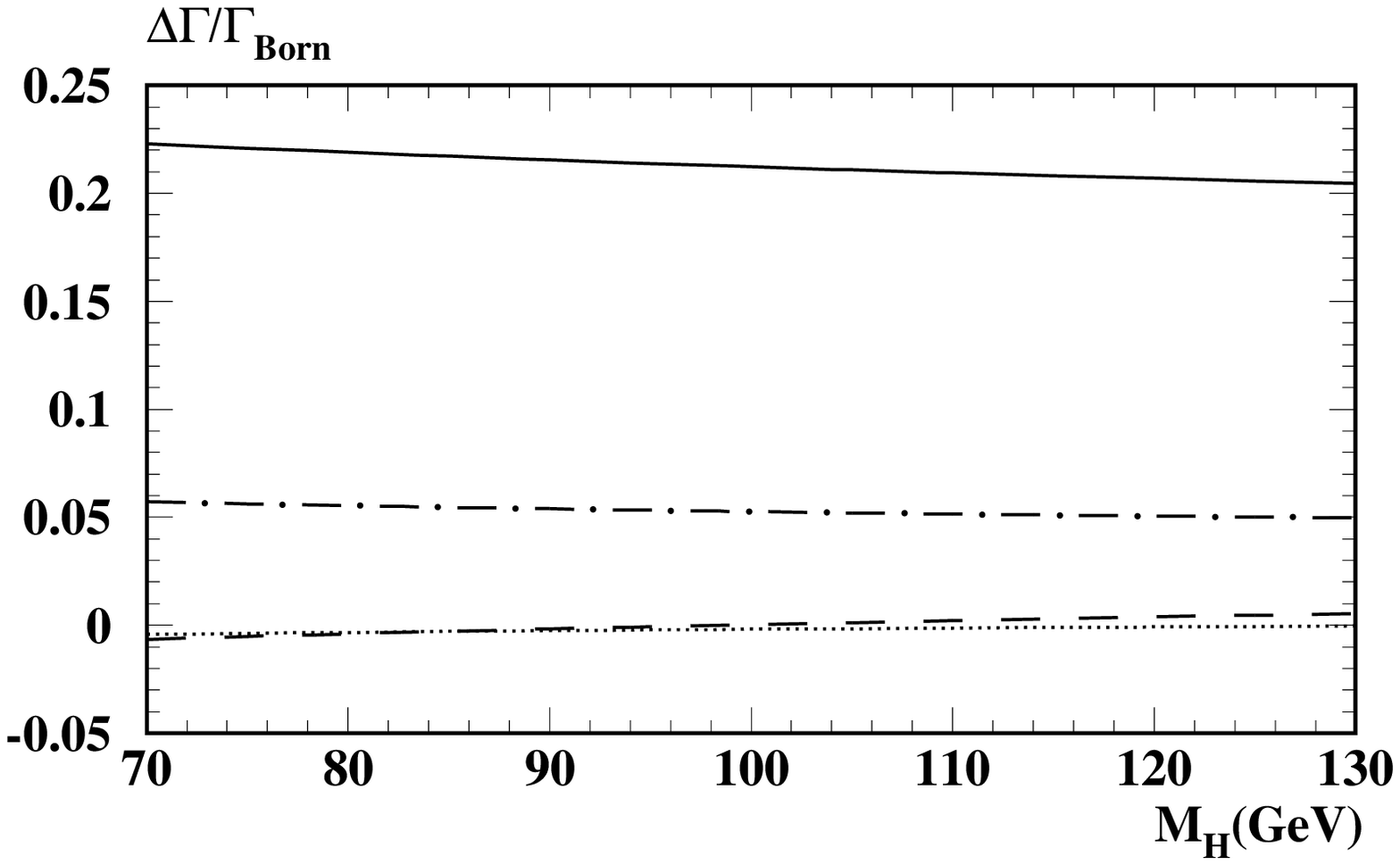}
 \end{tabular}

\vspace{3ex}
{\bf Figure \ref{fig5}}
 \end{center}
\end{figure}

\end{document}